\documentclass[twocolumn,prd,aps,superscriptaddress,showpacs,preprintnumbers]{revtex4}

\usepackage{amsmath}
\usepackage{verbatim}
\usepackage{graphicx}
\usepackage[usenames]{color}
\usepackage{psfrag}
\usepackage{epstopdf}
\usepackage{hyperref}

\def\beq{\begin{equation}}
\def\eeq{\end{equation}}
\def\bea{\setlength\arraycolsep{1.4pt}\begin{eqnarray}}
\def\eea{\end{eqnarray}}
\def\bit{\begin{itemize}}
\def\eit{\end{itemize}}

\begin{document}
\preprint{\tt TTK-10-01}

\title{Tight constraints on F- and D-term hybrid inflation scenarios}
\author{Richard Battye} \email{rbattye@jb.man.ac.uk}
\affiliation{Jodrell Bank Centre for Astrophysics, School of Physics and Astronomy, University of Manchester, Manchester, M13 9PL  UK}
\author{Bj\"orn Garbrecht} \email{garbrecht@physik.rwth-aachen.de}
\affiliation{Institut f\"ur Theoretische Physik, RWTH Aachen, 52056 Aachen, Germany}
\author{Adam Moss} \email{adammoss@phas.ubc.ca}
\affiliation{ Department of Physics \& Astronomy, University of British Columbia, Vancouver, BC, V6T 1Z1  Canada}

\date{\today}

\begin{abstract}
We use present cosmological data from the cosmic microwave background, large-scale structure and deuterium at high redshifts to constrain supersymmetric F- and D-term hybrid inflation scenarios including possible contributions to the CMB anisotropies from cosmic strings. Using two different realizations of the cosmic string spectrum, we find that the minimal version of the D-term model is ruled out at high significance. F-term models are also in tension with the data. We also discuss possible non-minimal variants of the models. 
\end{abstract}

\maketitle

{\em Introduction:} Present observations of the Universe provide strong evidence for a topology which is close to spatially flat and a close to scale invariant spectrum of density fluctuations. These are both predications of the simplest models of inflation based on the slow-roll approximation and hence provide prima-facie evidence for an inflationary epoch in the Early Universe. One of the basic features of inflation is that the potential needs to be very shallow in order to give rise to perturbations with the observed amplitude. This can be difficult to arrange in many particle physics motivated scenarios since it typically requires parametric fine-tuning in the theory or a trans-Planckian field-range for the inflaton.

One class of models which naturally circumvents this issue are the Supersymmetric (SUSY) Hybrid Inflation scenarios which have F- and D-term variants~\cite{hybrid,fterm,dterm}. In these models the tree level potential is completely flat and the slow-roll is induced by the radiative corrections. There are two key observational features of the minimal version of these scenarios:  a spectral index of density fluctuations, $n_{\rm S}$, which is 0.98--1.0, and cosmic strings are often produced during the phase transition which ends inflation~\cite{Jeannerot:2003qv}, for example, if the broken symmetry is $U(1)$.  For typical model parameters the amplitude of cosmic microwave background (CMB) anisotropies from strings will be $\sim 10\%$ of those produced by inflation~\cite{10per}.

This is particularly interesting since the data from the Wilkinson Microwave Anisotropy Probe (WMAP) 3 year release suggested that $n_{\rm S}=0.958\pm 0.016$ which appears to disfavor such models~\cite{wmap3}. In subsequent work we pointed out that this discrepancy could be ameliorated by the inclusion of a component of the CMB anisotropies from cosmic strings~\cite{Battye1}. In particular, we found that the minimal version of the F- and D-term scenarios, which have 6 free parameters, fitted the data as well as or better than the standard 6 parameter fit. Similar results, based on a different cosmic string spectrum, were reported by Bevis {\it et al}, although they did not specifically refer to the SUSY hybrid inflation models, rather they investigated a 6 parameter model with the string tension, $G\mu$, allowed to vary and $n_{\rm S}=1$ fixed~\cite{bevis}.

Since then cosmological data has moved on. WMAP have released their 5 year data~\cite{wmap5}, other CMB data on smaller angular scales has been published~\cite{othercmb}, the SDSS has been used to compute the matter power spectrum~\cite{sdss} and tight constraints have been derived on the baryon density, quantified by the value relative to the critical density, $\Omega_{\rm b}$, using Big Bang Nucleosynthesis (BBN) from measurements of deuterium at high redshift~\cite{BBN}. The last two of these are particularly interesting in this context since we found in our earlier analysis that models which fitted the data due to the inclusion a string component with $n_{\rm S}\approx 1$ typically have values of $h=H_0/100{\,\rm km}\,{\rm sec}^{-1}\,{\rm Mpc}^{-1}$ and $\Omega_{\rm b}h^2$ which are larger than the standard values~\cite{Battye1}.

In this {\it letter} we will revisit the constraints on the SUSY hybrid inflation models focusing on the minimal F- and D-term variants and on those that are extended by a possible additional mass-square term for the inflaton~\cite{BasShaKi,LythBoub,LinMcDonald,LindeLuestEtAl}. We will take into account the cosmic string component using two possible models. In model A we use the unconnected segment model~\cite{usm} (USM) to produce the string spectrum using an evolution model for the correlation length, rms string velocity and string wiggliness. This spectrum was used in our previous work~\cite{Battye1} and also in refs.~\cite{pog}. We also include results based on another spectrum created using the USM, which we will call model B. This spectrum, with different parameters from model A, was designed to match the results of the Abelian-Higgs simulations reported in ref.~\cite{bevis} (see ref.~\cite{Battye2} for a  discussion). It is our opinion that model A is likely to be closest  to the true cosmic string spectrum, but the two spectra span the different possibilities reported in the literature. We note that since the cosmic string spectrum will be $\sim 10\%$ of the total, the level of accuracy required is much less than for the adiabatic spectrum, and the two models give rise to qualitatively similar results.


{\em Methodology:} Our basic methods will be essentially the same as those described in ref.~\cite{Battye1}. The F-term superpotential is given by $W=\kappa{\hat S}({\hat{\bar G}}{\hat G}-M^2)$ where ${\hat S}$ is the singlet inflaton superfield and ${\hat G}$, ${\hat{\bar G}}$ are the waterfall fields responsible for the topological defects ~\cite{Jeannerot:2003qv}. We include the radiative corrections and minimal supergravity corrections, but ignore the curvature and tadpole terms. These terms were shown to have negligible effect in previous work, where we assumed a positive sign for the tadpole term. For the D-term model we use $W=\kappa{\hat S}{\hat{\bar G}}{\hat G}$. Spontaneous breaking of ${\rm U}(1)$ is induced by the D-term contribution $g^2(|G|^2-|{\bar G}|^2 + M^2)^2/8$ to the potential. Without supergravity corrections, the predictions for inflation and strings from the D-term model do not depend on the gauge coupling $g$. Taking account of minimal
supergravity corrections, we find
that for values of $g\stackrel{>}{{}_\sim}0.1$, the spectrum receives an additional blue-tilt, which corresponds to a less-good fit to the data~\cite{Battye1}. When compared to small  $g$, the hybrid inflation prediction for $n_{\rm S}$ lies substantially above the central value from the standard 6-parameter fit. In this work, we neglect minimal supergravity corrections, corresponding to a restriction of parameter space to
$g\stackrel{<}{{}_\sim}0.1$. Therefore in both the F- and D-term cases, the inflation model is specified by two parameters: the dimensionless coupling constant, $\kappa$, and the mass scale, $M$. It is possible to compute the three observable quantities $n_{\rm S}$, $G\mu$ and the scalar amplitude $\mathcal {P_{\rm R}}$ from these two parameters, and hence the fit is for 6 parameters (with the baryon density, $\Omega_{\rm b}h^2$, cold dark matter density $\Omega_{\rm c}h^2$, acoustic scale, $\theta_{\rm A}$ and optical depth, $\tau$, being the other 4). The reheat temperature is fixed to be $T_{\rm R}=10^{9}{\rm GeV}$ but allowing this to vary only has a minor effect on our results.

We refer to more details on the models to ref.~\cite{Battye1}, but in the present context, it is useful to recall the basic features distinguishing the F- and D-term models phenomenologically. For $\kappa^2/g^2<1/2$ the string  tension for the same value of $M$  is larger in the D-term model than the F-term. Hence, in order to comply with upper bounds on the string contribution to the CMB, D-term models predict  a smaller $M$, which also leads to a close to scale invariant spectrum, $n_{\rm S}\approx 1$.

We use COSMOMC~\cite{cosmomc} to perform the analysis. We include data  from WMAP5~\cite{wmap5}, other CMB experiments~\cite{othercmb}  and the SDSS~\cite{sdss}. In addition we will also include constraints on $\Omega_{\rm b}h^2$ from metal poor damped Lyman-$\alpha$ systems. These have been used to compute the deuterium abundance at $z\approx 2.6$. One finds that $\log(D/H)=-4.55\pm 0.03$, which implies that $\Omega_{\rm b}h^2=0.0213\pm 0.0010$~\cite{BBN}.

\begin{figure}
\centering
\mbox{\resizebox{0.45\textwidth}{!}{\includegraphics[angle=0]{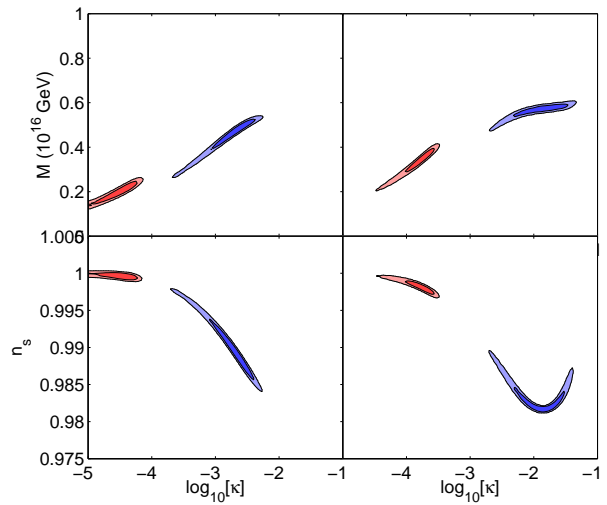}}}
\caption{\label{fig:results} 2D likelihoods for $D$-Term (red, smaller values of $\kappa$) and $F$-term (blue, larger values of $\kappa$) for string model A (left set of two) and B (right set of two) using CMB+SDSS data.}
\end{figure}

{\em Results:} We find that the 4 cosmological parameters are close to the values preferred by the standard 6 parameter fit. In Fig.~\ref{fig:results}, we present 2D likelihoods for various parameters of the F- and D-term models for cosmic string spectra A and B. The best-fitting values of $M$ and $\kappa$ are given in table~\ref{tab:mkappa}. We see that the values are not the same for the two string spectra.  The central values for model A have changed from those reported in ref.~\cite{Battye1} due to the inclusion of the new CMB and the SDSS data, although they remain within $\sim 1\sigma$.  The size of the errorbars have not changed significantly. The results for model B are somewhat different reflecting different ranges of $G\mu$ and $n_{\rm S}$ which are favoured using the different string spectra~\cite{Battye2}. Of particular importance is that model A restricts $n_{\rm S}=0.958\pm 0.013$ in a 7 parameter fit including $G\mu$ as the extra parameter, which is very similar to the 6 parameter case.  However,  $n_{\rm S}=0.970\pm 0.016$ for model B, which is more compatible with $n_{\rm S}=1$.

One can define the relative likelihood of two models, labelled 1 and 2, by $\Delta\chi^2=-2\log({\cal L}_1/{\cal L}_2)$. We will take model 1 to be the standard 6 parameter model using $n_{\rm S}$ and $A_{\rm S}$ to define the initial power spectrum. Positive values of this quantity indicate that model 2 fits the data better than the standard, whereas negative values indicate a less good fit. This quantity was used in ref.~\cite{wmap3} to quantify the level to which various variants of the standard cosmological model (for example, those with no reionization, or no dark matter) were poor fits to the data,  We present the computed values of $\Delta\chi^2$ for the F- and D-term models using the two cosmic string spectra in table~\ref{tab:delchi}.  We see that if only CMB data is included there is little to tell between the 6 parameter model and the F/D-term models, although the D-term model using string spectrum A is the least good fit. If  one includes more data from SDSS and BBN, typically the fit of the hybrid inflation models become less good.  

\begin{table}
\begin{center}
\begin{tabular}{|c|c|c|} \hline
 & $\log_{10} \kappa$ & $M /10^{16} {\rm GeV}$   \\ \hline
 $F$  (A) &  $-2.79 \pm 0.30$ & $0.450 \pm 0.062$ \\ \hline
 $F$  (B) & $-1.94 \pm 0.34$ & $0.563 \pm 0.038$ \\ \hline
 $D$  (A) &  $-4.55 \pm 0.21$  &  $0.196 \pm 0.030$ \\ \hline
 $D$  (B) & $-3.84 \pm 0.23$ & $0.330 \pm 0.047$ \\ \hline
\end{tabular}
\caption{ \label{tab:mkappa}
Constraints on minimal $F$ and $D$-term parameters using string models A and B.}
\end{center}
\vskip-.6cm
\end{table}

Concentrating on the case of CMB+SDSS+BBN, it is clear that the D-term model using both string spectra is a much worse fit than the 6 parameter model with $\Delta\chi^2=-17.6$ for string spectrum A and -12.2 for B. This shows that the D-term model is severely in tension with the data. This is because the best-fitting D-term models and those with $n_{\rm S}=1$, in the absence of the BBN prior, typically favour values of $\Omega_{\rm b}h^2\sim 0.024-0.025$. The situation is more complicated in the F-term case. In both cases the F-term model is disfavoured relative to the 6 parameter model, but for string spectrum B the $\Delta\chi^2$ is only $-5.6$. This would not be enough to exclude the model with high confidence, whereas for spectrum A the $\Delta\chi^2= -12.2$, which would. We have already alluded to the reason for this: spectrum B allows for larger values of $n_{\rm S}$ than the spectrum A.

In order to calibrate our interpretation of these values we have done two things. First, we computed $\Delta\chi^2$ for a 5 parameter model with $\tau=0$. For each combination of data, we find $\Delta\chi^2\approx -20$, with roughly $5 \sigma$ significance that $\tau \ne 0$ (for example, CMB data alone gives $\tau = 0.090 \pm 0.017$). We have also done the same for models with $n_{\rm S}=1$, that is a Harrison-Zel'dovich (HZ) spectrum. For CMB+SDSS data we find $\Delta \chi^2 = -11.8$, corresponding 
to the HZ model being excluded at $3.4\sigma$, and for CMB+SDSS+BBN data $\Delta\chi^2 = -16.8$, 
corresponding to $4.1\sigma$ ($n_{\rm S} = 0.951 \pm 0.012$). These values are presented in table~\ref{tab:delchi}. 

Comparing the $\Delta\chi^2$ for the F- and D-term models we can conclude that the D-term model is excluded at a level somewhere in the range of 3-4$\sigma$ even if strings are included, with the precise level dependent on which string spectrum is correct. Meanwhile the F-term model would be excluded at around $3 \sigma$ if string spectrum A is correct, but cannot be excluded is spectrum B is correct.

The statistical strength of our results is very strongly related to the value of $D/H$ measured in ref.~\cite{BBN} and the subsequent interpretation in terms of BBN. In order to explore the effects of systematic offsets in the value of $\Omega_{\rm b}h^2$ on our results we have computed the value of $\Delta\chi^2$ for a prior of $\omega_{\rm b}=\Omega_{\rm b}h^2={\bar\omega_{\rm b}}\pm 0.001$ (i.e. a different central value with the same statistical error as ref.~\cite{BBN}). For the D-term model with string spectrum A, the variation in $\chi^2$ for CMB+SDSS+BBN can be approximated by $ \Delta \chi^2  \approx 3160 \, {\bar\omega_{\rm b}} -  84.5$. Even if the central value were to increase by 0.002, the model would still be in tension with the data, $\Delta\chi^2 \approx -11$.

\begin{table}
\begin{center}
\begin{tabular}{|c|c|c|c|c|} \hline
& CMB & +SDSS & +BBN & +SDSS/BBN \\ \hline
$\tau=0$ & -21.7 & -20.4 & -20.8 & -20.2  \\ \hline
HZ & -9.4  & -11.8 & -14.6  & -16.8  \\ \hline
$F$  (A) & -1.2  & -4.4 & -10.2  & -13.2  \\ \hline
$F$  (B) & 1.2   & 0.2  & -4.0 &  -5.6 \\ \hline
$D$  (A) & -4.0  & -8.6 & -14.2 & -17.6  \\ \hline
$D$  (B) & 0.2 & -2.8  & -9.6 & -12.2 \\ \hline
\end{tabular}
\caption{ \label{tab:delchi}
$\Delta\chi^2=-2\log({\cal L}_1/{\cal L}_2)$ for minimal 6 parameter $F$ and $D$-term scenarios. These are defined as model 2; model 1 is the standard 6 parameter power law. We also present results relative to the 5 parameter $\tau=0$  and HZ models.}
\end{center}
\vskip-.6cm
\end{table}

{\em Non-minimal models:} Non-minimal F- and D-term models can be considered; these typically involve the inclusion of a negative mass-square term in the potential and can allow for values of $n_{\rm S}$ lower than 0.98 due to an enhanced negative curvature of the potential. In addition, at the point of horizon exit, the potential can be effectively flattened. This leads to a lower energy-scale of inflation and therefore to lower values for $M$ and a smaller string tension. In ref.~\cite{Battye1} we considered a particular class of F-term models~\cite{BasShaKi} characterized by the addition of the potential term $c_H^2 H^2 S^2$, where $H$ denotes the Hubble rate. This ``Hubble-dependent'' mass term generically arises in the presence of a non-minimal K\"ahler potential from the F-term that drives inflation.  

In the absence of F-terms, even for a non-minimal K\"ahler potential, no such additional mass-square correction can occur. This is why for the simplest realizations of D-term inflation, the exclusion of mass-square corrections is well-motivated. However, there are particular models where sub-dominant F-terms ({\it e.g.} from compactification in string models or from sneutrino potentials), that do not drive inflation, may be present~\cite{LythBoub,LinMcDonald,LindeLuestEtAl}. These can lead to mass-square corrections for the D-term potentials that may redden the spectrum and reduce the string contribution to the CMB. 

We aim to formulate the model in such a way that the number of parameters remains small ({\it e.g} no more than 7 parameters) and presume strings form. This can be achieved by the restriction to $g\stackrel{<}{{}_\sim}0.1$ and adding the potential term in the form of $g^2 c_H^2 S^2$. This way, the predictions remain independent of the gauge coupling $g$. We note that in presence of a negative mass-square term, there may also be models with $g\stackrel{>}{{}_\sim}0.1$, that are no longer disfavoured by the data. Unless one applies a restriction to more specific models, a systematic study of this region of parameter space would require additional free parameters in the analysis and is therefore not performed here.

\begin{figure}
\centering
\mbox{\hskip -0.5cm\resizebox{0.6\textwidth}{!}{\includegraphics[angle=0]{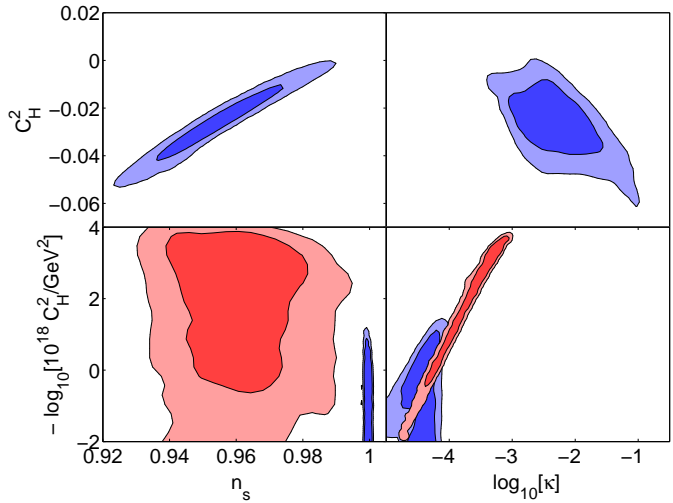}}}
\caption{\label{fig:nonmin} 2D likelihoods for non-minimal F- (top set of two) and  D-term (bottom set of two) for string model A. In the D-term case the two sets of contours show the two solutions admitted by the potential, as explained in the text.}
\end{figure}

In Fig.~\ref{fig:nonmin} we show 2D likelihoods for the non-minimal models using string spectrum A. In the F-term case, negative values of $c_H^2$ lead to an increased red-tilt in  $n_{\rm S}$, alleviating the tension with data present for $c_H^2=0$. The fit is slightly better than the 6 parameter model with $\Delta\chi^2 = 1.2$. We find marginalized values of $\log_{10} \kappa = - 2.28 \pm 0.46$, $M= \left(0.346  \pm 0.065 \right) \times 10^{16} \, {\rm GeV}$ and  $c_H^2= -0.027 \pm 0.011$. 

For the D-term model the situation is more complicated, since for fixed $\mathcal {P_{\rm R}}$  and $\kappa$ the potential can admit {\em two} solutions, each with different $M$ and parameters $n_{\rm S}$ and $G \mu$. In the analysis we treat these independently, running separate MCMC chains for each solution branch. The results of the analysis are shown in Fig.~\ref{fig:nonmin}. The solution for small $\kappa$ is essentially the same as the minimal case. Here, $M$ is higher and a non-zero $c_{H}^2 = 0$ does not effect the inflation dynamics. We find  $\log_{10} \kappa = - 4.45 \pm 0.22$ and $M= \left(0.203 \pm 0.032 \right) \times 10^{16} \, {\rm GeV}$, with $\Delta\chi^2 = -8.6$.

The other solution has a lower value of $M$, and a non-zero $c_{ H}^2$ flattens the potential near horizon exit. One gets sufficient e-foldings when  the inflaton is placed close to the top of the potential and the resulting $n_{\rm S}$ is redder, with the same slightly improved $\Delta\chi^2 = 1.2$ as in the F-term model. For this branch larger values of $\kappa$ are allowed, since the lower $M$ leads to a smaller string tension. We find  $\log_{10} \kappa = - 3.78 \pm 0.44$, $M= \left(0.116  \pm 0.071 \right) \times 10^{16} \, {\rm GeV}$ and  $-\log_{10} \left(10^{18} c_H^2 / {\rm GeV^2} \right) = 1.63 \pm 1.50$. We note that the upper bound on $\kappa$ is related to the upper bound on $G\mu$.
 

{\em Conclusions:} 
Minimal models of hybrid inflation rely only on two parameters ($\kappa$ and
$M$), and they do not suffer from the potential problem of trans-Planckian values of the inflaton field. Therefore they are predictive and it is possible to derive parametric constraints
from cosmological data. In Refs.~\cite{Battye1,bevis},
it has been found that a close to scale invariant spectrum of adiabatic perturbations
is in agreement with CMB data, when a sizable string component is present. In particular,
minimal models of hybrid inflation which predict these features appeared to be viable in that light,
a finding that may open interesting directions in model building~\cite{LindeLuestEtAl}.
Using recent CMB data, SDSS and measurements of the deuterium abundance leads us to a revison of this
picture: Minimal D-term models are ruled out at a level that can be estimated to be
$\approx 4\sigma$ (the same is true for the HZ spectrum)
and the tension between minimal F-term models and the data is increasing.
Non-minimal scenarios include the negative mass-square term models discussed here and also the  possibility of adding a negative-sign tadpole term~\cite{shafired}. This way, one can entirely remove tension with the data, but at the same time, also the very distinct predictions of minimal hybrid inflation for the range
of the spectral index are lost.


{\em Acknowledgments:} This research was supported by the Natural Sciences and Engineering Research Council of Canada and the British Columbia Knowledge Development Fund. We thank Douglas Scott for useful discussions.

\end{document}